\documentclass[tradiabstract]{aa}  

\def \Eqt{Eq.\thinspace}
\def \sect{Sect.\thinspace}
\def \fig{Fig.\thinspace}
\def \tab{Tab.\thinspace}

\newcommand{\RMd}{{\rm d}}

\def\eck#1{\left\lbrack #1 \right\rbrack}

\def\rund#1{\left( #1 \right)}

\def\ave#1{\left\langle #1 \right\rangle}

\def\N{{\cal N}}

\def\CM{{\tens{C}}}
\def\DM{{\tens{D}}}
\def\ZM{{\tens{Z}}}
\def\CMb{\boldsymbol{\tens{C}}}
\def\HMb{\boldsymbol{\tens{H}}}
\def\BM{{\tens{B}}}
\def\BMb{\boldsymbol{\tens{B}}}

\def\Real{{\rm I\mathchoice{\kern-0.70mm}{\kern-0.70mm}{\kern-0.65mm}%
  {\kern-0.50mm}R}}
\def\C{\rm C\kern-.42em\vrule width.03em height.58em depth-.02em
       \kern.4em}

\def\bx#1{\leavevmode\thinspace\hbox{\vrule\vtop{\vbox{\hrule\kern1pt
        \hbox{\vphantom{\tt/}\thinspace{\bf#1}\thinspace}}
      \kern1pt\hrule}\vrule}\thinspace}

\def\vc#1{{\mbox{\boldmath$#1$\unboldmath}}}
{\catcode`\@=11
\gdef\SchlangeUnter#1#2{\lower2pt\vbox{\baselineskip 0pt \lineskip0pt
  \ialign{$\m@th#1\hfil##\hfil$\crcr#2\crcr\sim\crcr}}}
}

\def\ueber#1#2{{\setbox0=\hbox{$#1$}%
  \setbox1=\hbox to\wd0{\hss$\scriptscriptstyle #2$\hss}%
  \offinterlineskip
  \vbox{\box1\kern0.4mm\box0}}{}}

\def\bx#1{\leavevmode\thinspace\hbox{\vrule\vtop{\vbox{\hrule\kern1pt
        \hbox{\vphantom{\tt/}\thinspace{\bf#1}\thinspace}}
      \kern1pt\hrule}\vrule}\thinspace}

{\catcode`\@=11
\gdef\SchlangeUnter#1#2{\lower2pt\vbox{\baselineskip 0pt \lineskip0pt
  \ialign{$\m@th#1\hfil##\hfil$\crcr#2\crcr\sim\crcr}}}
}
\usepackage{graphicx}
\usepackage[T1]{fontenc} %correct
\usepackage{lmodern} %correct
\usepackage{color}
 \usepackage{ucs}
\usepackage[utf8x]{inputenc}
\usepackage{url}
\usepackage{mathrsfs}
\usepackage{amsmath} %correct
\usepackage{amssymb}
\usepackage{appendix}
\usepackage{multirow}
\usepackage{graphicx} %correct
\usepackage{varioref}
 \usepackage[T1]{fontenc}
\usepackage{txfonts}
\usepackage{natbib}
\usepackage{subfigure}
\usepackage{enumerate}
\begin{document}
   \title{A New Data Compression Method and its Application to Cosmic Shear Analysis}

   \author{    Marika Asgari
    \inst{1}
          \&
          Peter Schneider
          \inst{2}
          }

   \institute{SUPA, Institute for Astronomy, University of Edinburgh, Royal Observatory, Blackford Hill, Edinburgh, EH9 3HJ, U.K.
             \email{ma@roe.ac.uk}       
        \and
        Argelander-Institut f\"ur Astronomie, Bonn University             
             }

%    \date{Received September 15, 1996; accepted March 16, 1997}

\abstract
{Future large scale cosmological surveys will provide huge data sets
  whose analysis requires efficient data compression. In particular,
  the calculation of accurate covariances is extremely challenging
  with increasing number of statistics used.}  {The aim of the present
  work is to introduce a formalism for achieving efficient data
  compression, based on a local expansion of statistical measures
  around a fiducial cosmological model. We specifically apply and test
  this approach for the case of cosmic shear statistics. In addition,
  we study how well band powers can be obtained from measuring shear
  correlation functions over a finite interval of separations.}  {We
  demonstrate the performance of our approach, using a Fisher analysis
  on cosmic shear tomography described in terms of E-/B-mode
  separating statistics (COSEBIs).}  {We show that our data
  compression is highly effective in extracting essentially the full
  cosmological information from a strongly reduced number of
  observables. Specifically, the number of statistics needed decreases
  by at least one order of magnitude relative to the COSEBIs, which
  already compress the data substantially compared to the shear
  two-point correlation functions.  The efficiency appears to be
  affected only slightly if a highly inaccurate covariance is used for
  defining the compressed statistics, showing the robustness of the
  method. Furthermore, we show the strong limitations on the
  possibility to construct top-hat filters in Fourier space, for which
  the real-space analog has a finite support, yielding strong bounds
  on the accuracy of band power estimates.}  {We conclude that an
  efficient data compression is achievable and that the number of
  compressed statistics depends on the number of model parameters.
  Furthermore, a band convergence power spectrum inferred from a
  finite angular range cannot be accurately estimated.  The error on
  an estimated band-power is larger for a narrower filter and a
  smaller angular range which for relevant cases can be as large as
  10\%.}


   \maketitle
%
%________________________________________________________________

\section{Introduction}

Future cosmological surveys are faced with the difficulty 
to extract cosmological parameters from their wealth of
observables. Taking Euclid\footnote{http://sci.esa.int/euclid/,
  \cite{EuclidRB}} as example, statistics to be obtained from 
the data include second-order shear statistics across
several populations of source galaxies, which -- using the common
usage -- will be termed `redshift bins' throughout this paper.
As shown in \cite{2010SEK} and \cite{2012ASS},
 the COSEBIs (Complete Orthogonal E-/B-mode Integrals) form 
appropriate combinations of the shear two-point correlation
functions $\xi_\pm(\theta)$ which cleanly separate E- and B-mode
shear \citep[see, e.g.,][]{Critt02,SvWM02}. In addition, COSEBIs are
highly efficient in terms of data compression, since essentially all
cosmological information is contained in a small number of COSEBIs 
\cite[see, e.g.,][for applications of COSEBIs to cosmic shear data
sets]{Kilbinger13,HEH14}.  

The efficiency of data compression decreases, however, if several
populations of sources are used. For example,
with $\sim 10$ redshift bins, the total number
of COSEBIs which should be used to extract cosmological information is
of order 500. Furthermore,
higher-order shear information contains additional,
valuable information -- both regarding cosmological parameters as well
as for calibrating the shear data --
and should be taken into account. Since third-order
shear statistics depends on three variables (say, three sides of a 
triangle), and combinations of three redshift
bins, the number of observables for third-order shear
statistics which needs to be considered is almost certainly
considerably larger than that for second-order shear statistics. 
Furthermore, shear-peak statistics has been shown to yield powerful
constraints and should likewise be considered \citep[see, e.g.,][and
references therein]{Marian13}. Therefore, the number
of pure shear observables will be several thousands, although the
number of cosmological parameters to be determined is of order a
dozen. 

In practice, issues are even more complicated, in that astrophysical
and other systematics need to be accounted for. For example, effects
of intrinsic alignments \citep[see, e.g.,][and references therein]{JoBr10}
need to be mitigated, by including further
observables, i.e., the galaxy-galaxy lensing signal and the galaxy
correlation functions. Even if one uses a COSEBI-like data compression
for them (e.g., \citealt{2013arXiv1302.2401E}), the number of redshift
combinations will still lead to a strongly enhanced number of
observables. 

One of the major difficulties in analyzing this data is the
determination of the expectation values for these observables as a
function of the parameters and, in particular, the estimation of their
covariance matrix.  If one determines the covariance as a sample
variance of different numerical realizations, one needs many more
realizations than the dimension of the data vector in order to get a
reliable estimate of the covariance matrix and its inverse (see, e.g.,
\citealt{2007A&A...464..399H}).  Because of this difficulty, data
compression is mandatory for any analysis of survey data.

In this paper, we suggest a form of data compression that is based on
the sensitivity of the various observables to the parameters that are
to be estimated. The cosmological parameters currently are, and
until the launch of Euclid will be even more, strongly constrained,
and thus only a relatively small volume in parameter space needs to be
explored.\footnote{Additional parameters, needed to parametrize intrinsic
alignment effects, may be less well constrained from independent data
sets or theoretical models.} We will therefore assume that the
relevant parameter region is small, which allows us to define linear
combinations of observables based on a low-order Taylor expansion of
the dependence of these observables on parameters,
which should contain almost all the
cosmological information in the data.

In the following section we introduce our data compression formalism
for general observables (statistics). We then specialize this method
in \sect\ref{sectAppCOSEBIs} to study how this strategy works for
COSEBIs compression. In \sect\ref{SectCosmology} we specify our
cosmological model which will be used for the results section. In
\sect\ref{SectResults} we first illustrate the weight functions for
the compressed statistics made of COSEBIs, then using a Fisher
formalism we explore the efficiency of the compressed versus regular
COSEBIs. Section\thinspace\ref{SectBandPower} is dedicated to
mimicking a band power spectrum using linear combinations of COSEBIs.
Finally we conclude in \sect\ref{SectConclusions}.

\section{Formalism}
\label{secFormalism}
Let $\hat X_n$ be the statistics obtained from the data, $1\le n\le N$,
with expectation value $\ave{\hat X_n}=X_n(\phi_\mu)$, where the
$\phi_\mu$, $1\le \mu\le P$, denotes the parameters of the model,
including the cosmological parameters as well as others.
Assuming that the uncertainty in the parameters is `small', we
consider an expansion of the functions $X_n(\phi_\mu)$ around the
fiducial value $X_n^{\rm f}=X_n(\phi_\mu^{\rm f})$,
\begin{equation}
\label{eq1}
 X_n(\phi_\mu)=X_n^{\rm f}+\DM_{n\mu}p_\mu+{1\over 2}\ZM_{n\mu\nu}p_\mu
p_\nu \;,
\end{equation}
where $p_\mu=\phi_\mu-\phi_\mu^{\rm f}$, and 
\begin{equation}
\label{eq2}
{\DM}_{n\mu}=\rund{\partial X_n \over \partial
  \phi_\mu}_{|\phi_\kappa^{\rm f}} \; ; \quad
{\ZM}_{n\mu\nu}=\rund{\partial^2 X_n \over \partial
  \phi_\mu\, \partial \phi_\nu}_{|\phi_\kappa^{\rm f}}
\end{equation}
are the first and second derivatives of the expectation values with
respect to the model parameters, taken at the fiducial point in
parameter space. Here and below, summation over repeated indices is
implied, unless noted otherwise. 

We assume that the likelihood ${\cal L}(\chi^2)$ is a monotonically
decreasing function of 
\begin{equation}
\label{eq3}
\chi^2=\eck{ \hat X_m-X_m(\phi_\mu)} \rund{\CM^{-1}}_{mn}
\eck{ \hat X_n-X_n(\phi_\mu)} \;,
\end{equation}
where $\CMb$ is the covariance matrix of the observables $\hat X_n$.
Maximizing the likelihood then requires to find the minimum of
$\chi^2$ with respect to the parameters; using \Eqt\eqref{eq1}, we obtain
\begin{align}
\label{eq4}
\partial \chi^2\over \partial p_\kappa &=-2\eck{\DM_{m\kappa}+\ZM_{m\kappa\mu} p_\mu}\rund{\CM^{-1}}_{mn} \nonumber \\
&\times\eck{\hat X_n-X_n^{\rm f}-\DM_{n\mu}p_\mu-{1\over 2}\ZM_{n\mu\nu}p_\mu p_\nu}=0 \;.
\end{align}
In this equation, we have neglected the dependence of the covariance
matrix on the parameters, either because $\CMb$ is determined from the
data itself, or because the dependence of $\CMb$ on the parameters is
assumed to be weak.  
From \Eqt\eqref{eq4}, we see that the determination
of the parameters $p_\mu$ does involve the observables $\hat X_n$ only
in the linear combinations
\begin{equation}
\label{eq5}
\hat F_\kappa := \DM_{m\kappa}\rund{\CM^{-1}}_{mn} \hat X_n \; ;\quad
\hat S_{\kappa\nu}:=\ZM_{m\kappa\nu}\rund{\CM^{-1}}_{mn}\hat X_n\;,
\end{equation}
with expectation value
\begin{equation}
\label{eq6}
F_\kappa := \DM_{m\kappa}\rund{\CM^{-1}}_{mn} X_n \; ;\quad
S_{\kappa\nu}:=\ZM_{m\kappa\nu}\rund{\CM^{-1}}_{mn} X_n\;.
\end{equation}
Thus, the expansion of the expectation values of the original
observables $\hat X_n$ around a fiducial model motivates the
definition of linear combinations of observables which contain all the
information about the parameters $\phi_\kappa$, provided the second-order
expansion is accurate. The set (5) of $P+P(P+1)/2=P(P+3)/2$
observables thus is expected to allow for an efficient data
compression (note that $\hat S_{\mu\nu}=\hat S_{\nu\mu}$).

In order to obtain the new observables $\hat F_\kappa$ and $\hat
S_{\kappa\nu}$, one first needs to estimate the covariance $\CMb$ of
the original observables which, due to the high dimensionality in
future cosmological surveys, provides a real challenge. However, the
covariance $\CMb$ is needed here for the definition of appropriate
combinations of observables, and not for parameter estimates. Hence,
an approximation for $\CMb$ may be expected to be sufficient for this
purpose.  Disregarding the parameter dependence of $\CMb$ in the
derivation of \Eqt\eqref{eq4} provides such an approximation which
avoids the necessity to obtain a large covariance matrix for more than
one cosmological model.  If the approximation for $\CMb$ deviates
substantially from the true covariance, we expect that the new
observables do not contain the full information about the parameters,
since they deviate from the `optimal' combination of the original
$\hat X_n$.  Hence, the better the initial estimate of $\CMb$, the
more efficient the new observables will be.

Thus, we propose a strategy to first obtain an approximation for the
covariance $\CMb$, based on which the new observables $\hat F_\kappa$
and $\hat S_{\kappa\nu}$ are defined. The number of these observables
is substantially smaller than the original ones, and hence an accurate
estimation of their covariance can be obtained from fewer simulations
compared to $\CMb$. On the other hand, the number of new observables
is substantially larger than the number of parameters, which is
expected to provide a mitigation for the choice of non-optimal
combinations from an approximate form of $\CMb$. It is for this reason
that we consider the second-order derivatives of the original
observables; the first-order ones coincide with that of the 
Karhunen--Lo\`eve method for the case of known covariance \citep[see,
e.g.,][]{1997ApJ...480...22T}. 

We now combine the new observables $\hat F_\kappa$ and $\hat
S_{\kappa\nu}$ into the $N'=P(P+3)/2$ compressed quantities $\hat X^{\rm c}_i$. According to \Eqt\eqref{eq5}, we can write
\begin{equation}
\label{eq7}
\hat {\vc X^{\rm c}}=\HMb \CMb^{-1} \hat{\vc X}\equiv \BMb \hat{\vc X} \;,
\end{equation}
where we use vectorial notation for the $\hat { X}^{\rm c}_i$ and
$\hat X_n$. The $N'\times N$ (rows $\times$ columns) matrix $\HMb$ is
given in terms of first and second partial derivatives of the
functions $X_n(\phi_\kappa)$ at the fiducial point in parameter space
and $\BMb=\HMb \CMb^{-1}$ is the compression matrix.  Accordingly, the
covariance matrix of $\hat{\vc X^{\rm c}}$ is given as
\begin{equation}
\label{eq8}
 \CMb^{\rm c}=\BMb\CMb \BMb^{\rm t}\;,
\end{equation}
where the superscript `t' denotes the transpose of a matrix.
The $\chi^2$-function in terms of the new
observables is
\begin{equation}
\label{eq9}
\chi^2=\eck{\hat{\vc X^{\rm c}}-{\vc X^{\rm c}}}^{\rm t}
{ (\CMb^{\rm c})^{-1}}
\eck{\hat{\vc X^{\rm c}}-{\vc X^{\rm c}}} \;.
\end{equation}
From what was discussed above, the covariance $\CMb^{\rm c}$ should be
calculated from $\CMb$ only if an accurate estimate of the latter can
be obtained; in general, it will be much more practical to determine
$\CMb^{\rm c}$ directly, e.g., from simulations.

Provided that $\CMb$ can be determined accurately, we can solve
\Eqt\eqref{eq4} for the parameters $\phi_\mu=p_\mu+\phi^{\rm f}_\mu$. Writing it in terms of the
new observables, \Eqt\eqref{eq4} becomes
\begin{equation}
\label{eq10}
\Delta F_\kappa+\Delta S_{\kappa\nu}\,p_\nu
=\rund{\DM_{m\kappa}+\ZM_{m\kappa\nu} p_\nu}  \rund{\CM^{-1}}_{mn}
\rund{\DM_{n\nu}p_\nu+{1\over 2}\ZM_{n\mu\nu} p_\mu p_\nu}\;,
\end{equation}
with $\Delta F_\kappa=\hat F_\kappa-F_\kappa^{\rm f}$, $\Delta
S_{\kappa\nu} = \hat S_{\kappa\nu}-S_{\kappa\nu}^{\rm f}$. If we then
expand $p_\mu=p_\mu^{(1)}+p_\mu^{(2)}$, where $p_\mu^{(1)}$
($p_\mu^{(2)} $) is first (second) order in the $\Delta F_\kappa$,
$\Delta S_{\kappa\nu}$, we obtain to first order
\begin{equation}
\label{eq11}
\Delta F_\kappa = \DM_{m\kappa} \rund{\CM^{-1}}_{mn} \DM_{n\nu}p_\nu^{(1)}
\equiv U_{\kappa\nu} p_\nu^{(1)}\;,
\end{equation}
from which we can easily obtain $p_\nu^{(1)}$ from the inverse of the
symmetric matrix $\boldsymbol{\tens{U}}$,
$p_\nu^{(1)}=\rund{\tens{U}^{-1}}_{\nu\mu}\,\Delta F_\mu$. The
second-order terms lead to the equation
\begin{equation}
\label{eq12}
\Delta S_{\kappa\nu}\,p_\nu^{(1)}={\tens U}_{\kappa\nu}\,p_\nu^{(2)}
+\rund{{1\over
    2}{\tens G}_{\kappa\mu\nu}+{\tens G}_{\mu\kappa\nu}}p_\mu^{(1)}p_\nu^{(1)} \;,
\end{equation}
where we defined 
\begin{equation}
\label{eq13}
{\tens G}_{\kappa\mu\nu}:=\DM_{m\kappa}\rund{\CM^{-1}}_{mn} \ZM_{n\mu\nu}\;.
\end{equation}
With the foregoing solution for $p_\mu^{(1)}$ and the inverse of $\boldsymbol{\tens U}$,
this can be immediately solved for $p_\mu^{(2)}$.

\section{Application to COSEBIs}
\label{sectAppCOSEBIs}
We will now apply the method of the previous section to a specific
statistics for cosmic shear measurements, the COSEBIs (see
\citealt{2010SEK}).  They provide a complete representation of the
shear two-point correlation functions (2PCFs) in a given finite
interval of angular scales, chosen such that they cleanly separate
between E- and B-modes \citep{Critt02,SvWM02}.  In our previous work
(\citealt{2012ASS}) 
we showed that COSEBIs also provide an efficient means of data
compression, since the full cosmological information contained in the
2PCFs can be recovered with a small number of COSEBIs.  However, in
the case of several redshift bins for the source galaxies, the number
of components grows with the number of tomographic redshift bins, $r$,
by a factor of $r(r+1)/2$.  In this section we use the formalism
explained in \sect\ref{secFormalism} to obtain a way to compress the
number of relevant statistical quantities and compare the results with
a full COSEBIs analysis.  The E-mode COSEBIs are related to the 2PCFs
via
\begin{equation}
\label{EqEnofTn}
 E_n^{(ij)} = \frac{1}{2} \int_{\theta_{\rm min}}^{\theta_{\rm
     max}}\RMd\vartheta\,\vartheta\: 
 [T_{+n}(\vartheta)\,\xi^{(ij)}_+(\vartheta) +
 T_{-n}(\vartheta)\,\xi^{(ij)}_-(\vartheta)]\;, 
\end{equation} 
where $1\le i,j\le r$ label the redshift bins considered.
The COSEBIs are defined for a given range of angular separations,
$[\theta_{\rm min},\theta_{\rm max}]$, i.e., the $T_{\pm
  n}(\vartheta)$ are zero outside this interval. They form a complete
basis for all filter functions that are defined on a finite angular
range and satisfy the conditions
\begin{equation}
\label{EqTCond}
\int_{\theta_{\mathrm{min}}}^{\theta_{\mathrm{max}}}
\mathrm{d}\vartheta\,\vartheta\,T_+(\vartheta) 
=0= \int_{\theta_{\mathrm{min}}}^{\theta_{\mathrm{max}}}
\mathrm{d}\vartheta\,\vartheta^3\,T_+(\vartheta)\;,
\end{equation}
which are the necessary and sufficient conditions for separating the E-
and B-modes obtained from the shear two-point correlation function
measured on a finite interval and for
removing ambiguous E-/B-modes \citep{2007A&A...462..841S}. 
As a result any allowed filter function is a linear combination of them.
E-mode COSEBIs are related to the power spectrum by 
\begin{equation}
\label{EqEnofWn}
E_n^{(ij)} = \int_0^{\infty}
\frac{\RMd\ell\,\ell}{2\pi}P^{(ij)}_{\mathrm{E}}(\ell)W_n(\ell)\;, 
\end{equation} 
where $P^{(ij)}_\mathrm{E}$ is the E-mode convergence cross-power
spectra of redshift bins $i$ and $j$ and
\begin{equation}
 W_n(\ell)=\int_{\theta_{\mathrm{min}}}^{\theta_{\mathrm{max}}} \mathrm{d}\vartheta\,\vartheta\,T_+(\vartheta)\, {\rm J}_0(\vartheta\ell)\;,
\end{equation}
where ${\rm J}_0$ is the zeroth-order Bessel function of the first
kind (see \citealt{2010SEK} and \citealt{2012ASS} where the filters
are defined and shown).

In the following, we use the logarithmic COSEBIs which yield a more
efficient data compression than the linear COSEBIs.  The Log-COSEBIs
$T_{+n}(\vartheta)$ filters are polynomials in $\ln(\vartheta)$ (see
\citealt{2010SEK}), i.e., they have more oscillations at small scales
and hence are more sensitive to variations of the shear 2PCFs on those
scales. As it turned out, an approximately uniform distribution of
roots of the weight function in logarithmic angular scales 
covers the cosmological information in the shear 2PCFs with a smaller
number of components.

In order to apply the method of the past section for obtaining a
compressed version of COSEBIs, we need to find their (approximate)
covariance matrix 
for a given cosmology, in addition to their first- and second-order
derivatives with respect to the cosmological parameters.  The new set
of statistics are related to the COSEBIs via the compression matrix,
$\BMb$, defined before in \Eqt\eqref{eq7}, 
\begin{equation}
\label{EqBtoE}
 E_\mu^{\rm c}= \BM_{\mu \N} E_{\N}=\BM_{\mu nij} E^{(ij)}_{n}\;,
\end{equation} 
where the new index 
\begin{equation}
\label{EqN}
 \N=\left[(i-1)r-\frac{(i-1)(i-2)}{2}+(j-1)\right]n_{\rm max}+n\;
\end{equation}
is a combination of the three indices $i,j$ and $n$, $n_{\rm max}$ is the
maximum order of COSEBIs considered, and $r$ is the total number of redshift bins.  

\section{Cosmological Model, Survey Parameters and Covariance}
\label{SectCosmology}
\begin{table}
\caption{\small{The fiducial cosmological parameters consistent with the
    WMAP 7-years results, and the underlying true parameters consistent with Planck.
The normalization of the power spectrum, $\sigma_8$, is the standard
deviation of perturbations in a sphere of radius $8 h^{-1} {\rm Mpc}$ today. 
$\Omega_\mathrm{m}$, $\Omega_\Lambda$, and $\Omega_\mathrm{b}$ are the matter, 
the dark energy and the baryonic matter density parameters, respectively.
$w_0$ is the dark energy equation of state parameter, which is equal to the ratio of dark energy pressure to its density. 
The spectral index, $n_\mathrm{s}$, is the power of the initial power spectrum. 
The dimensionless Hubble parameter, $h$, characterizes the rate of expansion today. }}
\begin{center}
\begin{tabular}{  c | c | c | c | c | c | c | c | }
  \cline{2-8}
  & $\sigma_8$ & $\Omega_\mathrm{m}$ & $\Omega_\Lambda$ & $w_0$ & $n_\mathrm{s}$ & $h$ & $\Omega_\mathrm{b}$ \\
  \hline
  \multicolumn{1}{|c|}{\multirow{1}{*}{Fiducial}} & 0.8  & 0.27  & 0.73  &  $-1.0$ &  0.97 & 0.70 & 0.045 \\
  \hline
  \multicolumn{1}{|c|}{\multirow{1}{*}{True}} & 0.83  & 0.31  & 0.68  &  $-1.1$ &  0.96 & 0.67 & 0.049 \\
  \hline
\end{tabular}
\end{center}
\label{Tabcosmparam}
\end{table}

A cold dark matter (CDM) cosmological models with a dynamical dark
energy, characterized by its equation-of-state parameter, $w_0$, is
used throughout this work (for references to $w$CDM models, see
\citealt{2003RvMP...75..559P}, and references therein).
\tab\ref{Tabcosmparam} contains the two sets of parameter values
considered here.  The fiducial model is used for obtaining the
compressed COSEBIs (CCOSEBIs hereafter), while the assumed `true'
underlying cosmology is slightly different.  That means, we calculate
the CCOSEBIs according to the Eqs. of Sect.2, using the covariance and
parameter derivatives of the COSEBIs, $\CMb$, $\boldsymbol{\DM}$ and
$\boldsymbol{\ZM}$, for the fiducial cosmology, but these new
observables $E_\mu^{\rm c}$ are applied using the `true' cosmological
model.  The linear matter power spectrum is calculated using the
\cite{1984ApJ...285L..45B} transfer function and a primordial
power-law power spectrum with spectral index $n_s$.  For non-linear
scales, the halo fit formula of \cite{2003MNRAS.341.1311S} is adopted.

For a cosmic shear analysis, we need the survey parameters and the
redshift distribution of the galaxies. The latter is 
characterized by \citep[see, e.g.,][]{1996ApJ...466..623B}
\begin{equation}
p(z)\propto\left(\frac{z}{z_0}\right)^\alpha \exp\left[-\left(\frac{z}{z_0}\right)^\beta\right]\;,
\end{equation}
for $z_{\rm min}\leq z\leq z_{\rm max}$ where the parameters,
$\alpha$, $\beta$, $z_0$, $z_{\rm min}$ and $z_{\rm max}$ depend on
the survey. \tab\ref{TabSurvey} summarizes the survey and redshift
parameters assumed in our analysis.

\begin{table}
\caption{\small{The parameters of a fiducial large future survey. 
$\alpha$, $\beta$, and $z_0$ characterize the total redshift
distribution of sources,  
while $z_\mathrm{min}$ and $z_\mathrm{max}$ indicate the minimum and
the maximum redshifts of the sources use for the cosmic shear analysis. 
Here $A$ is the survey area in units of deg$^2$, $\sigma_{\epsilon}$
is the galaxy intrinsic ellipticity dispersion,  
and $\bar n$ is the mean number density of sources per square
arcminute in the field.}} 
\begin{center}
\begin{tabular}{|c|c|c|c|c|c|c|c|}
\cline{1-8}
  \multicolumn{5}{|c|}{z-distribution parameters} 
  &\multicolumn{3}{|c|}{survey parameters}
\\ \cline{1-8}
  $\alpha$ & $\beta$ & $z_0$ & $z_\mathrm{min}$ & $z_\mathrm{max}$ & $A$ & $\sigma_{\epsilon}$ & $\bar n$\\ \cline{1-8}
 2.0 &  1.5 & 0.71 & 0.0 & 2.0 & 20000 & 0.3 & 35 \\ \cline{1-8}
\end{tabular}
\end{center}
\label{TabSurvey}
\end{table}

We assume Gaussian shear fields to find the covariances needed for
obtaining CCOSEBIs and also for the Fisher analysis (see
\citealt{2008A&A...477...43J}). The relation between the E-COSEBIs
covariance and the convergence power spectrum for redshift bin pairs
$ij$ and $kl$ is
\begin{align}
\label{EqCov}
\CM_{mn}^{(ij)(kl)} & \equiv  \langle E^{(ij)}_m E^{(kl)}_n\rangle
-\langle E^{(ij)}_m\rangle\langle E^{(kl)}_n\rangle \nonumber \\
& = \frac{1}{2 \pi A}\int_0^{\infty} 
\mathrm{d}\ell\:\ell\:W_m(\ell)W_n(\ell)\nonumber \\
&\times \left(
  \bar{P}^{(ik)}_{\mathrm{E}}(\ell)\bar{P}^{(jl)}_{\mathrm{E}}(\ell)
+\bar{P}^{(il)}_{\mathrm{E}}(\ell)\bar{P}^{(jk)}_{\mathrm{E}}(\ell)\right)\;,
\end{align} 
where 
\begin{equation}
\label{EqP}
\bar{P}^{(ik)}_{\mathrm{E}}(\ell):= P^{(ik)}_{\mathrm{E}}(\ell)
+\delta_{ik}\frac{\sigma_{\epsilon}^2}{2\bar{n}_i}\;,
\end{equation} 
$A$ is the survey area, $\sigma_\epsilon$ is the galaxy intrinsic
ellipticity dispersion and $\bar{n}_i$ is the average galaxy number
density in redshift bin $i$. The overall shape of the COSEBIs
covariance is shown in \cite{2012ASS}.

\section{Results}
\label{SectResults}
This section is dedicated to our results. The filter functions of the
CCOSEBIs for the fiducial cosmology are shown first,
followed by a figure-of-merit analysis. We
compare the figure-of-merit values for cases where the covariance is
known versus the use of a wrong covariance in constructing the CCOSEBIs.

\subsection{Weight functions of Compact COSEBIs}

Inserting Eqs.\thinspace\eqref{EqEnofTn} and \eqref{EqEnofWn} into
\Eqt\eqref{EqBtoE} results in relations between $\vc E^{\rm c}$ and the
COSEBIs filters,
\begin{equation}
 E^{\rm c}_\mu= \frac{1}{2} \int_{\theta_{\rm min}}^{\theta_{\rm max}}
\RMd\vartheta\,\vartheta
 [\BM_{\mu n ij}T_{+n}(\vartheta)\,\xi^{(ij)}_+(\vartheta) 
+ \BM_{\mu nij}T_{-n}(\vartheta)\,\xi^{(ij)}_-(\vartheta)]\;,
\end{equation} 
and
\begin{equation}
 E^{\rm c}_\mu= \int_0^{\infty} \frac{\RMd\ell\,\ell}{2\pi}
\BM_{\mu nij} W_n(\ell)\,P^{(ij)}_{\mathrm{E}}(\ell)\;.
\end{equation}
For each redshift bin pair, $ij$, a set of $N'=P(P+3)/2$ ($P$ is the
number of free parameters) filters exist. The new filters in real and
Fourier space, respectively, are 
\begin{align}
\label{eqWT}
T^{\rm c}_{\pm \mu ij}(\vartheta)&=\BM_{\mu nij}T_{\pm n}(\vartheta)\\ \nonumber
W^{\rm c}_{\mu ij}(\ell)&=\BM_{\mu nij}W_n(\ell)\;.
\end{align}
With the above definitions we can rewrite the compressed statistics,
$\vc E^{\rm c}$, in terms of the compressed filter functions,
\begin{equation}
 E^{\rm c}_\mu= \frac{1}{2} \int_{\theta_{\rm min}}^{\theta_{\rm max}}
\RMd\vartheta\,\vartheta
 [T^{\rm c}_{+ \mu ij}\,\xi^{(ij)}_+(\vartheta) 
+ T^{\rm c}_{- \mu ij}\,\xi^{(ij)}_-(\vartheta)]\;,
\end{equation} 
and
\begin{equation}
 E^{\rm c}_\mu= \int_0^{\infty} \frac{\RMd\ell\:\ell\:}{2\pi}
W^{\rm c}_{\mu ij}(\ell)\,P^{(ij)}_{\mathrm{E}}(\ell)\;.
\end{equation}
Multiplying each $E^{\rm c}_\mu$ by a constant has no effect on the
information level. We can therefore normalize the filter functions for
each compressed statistic separately, so that
\begin{align}
\label{eqNormalization}
 \sum_{ij}\frac{1}{\Delta \vartheta}\int_{\theta_{\rm min}}^{\theta_{\rm max}}\RMd\vartheta\: 
 [T^{\rm c}_{+ \mu ij}(\vartheta)]^2=1\;.
\end{align}
\begin{figure*}
  \begin{center}
    \begin{tabular}{c}
      \resizebox{180mm}{!}{\includegraphics{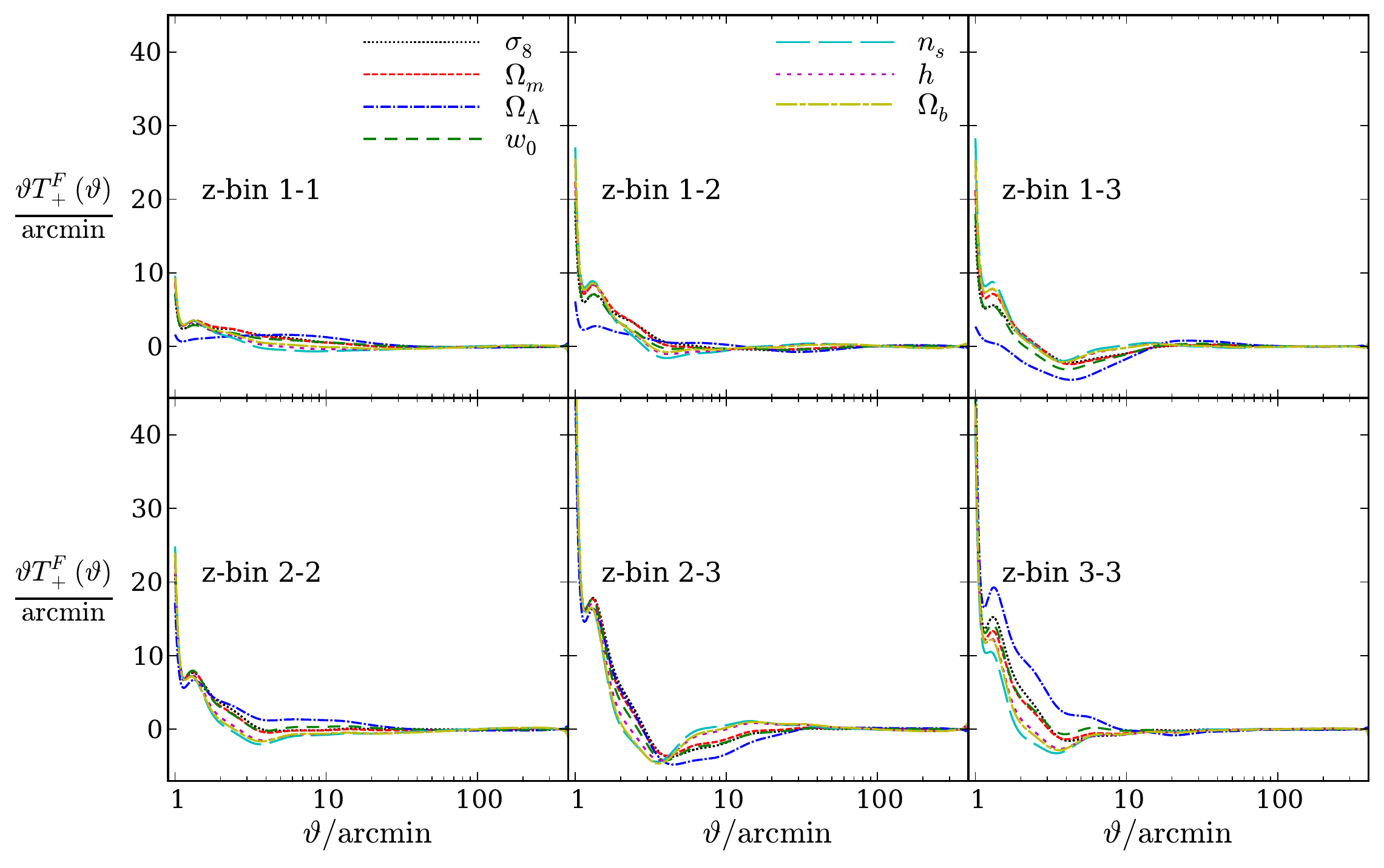}} \\
      \resizebox{180mm}{!}{\includegraphics{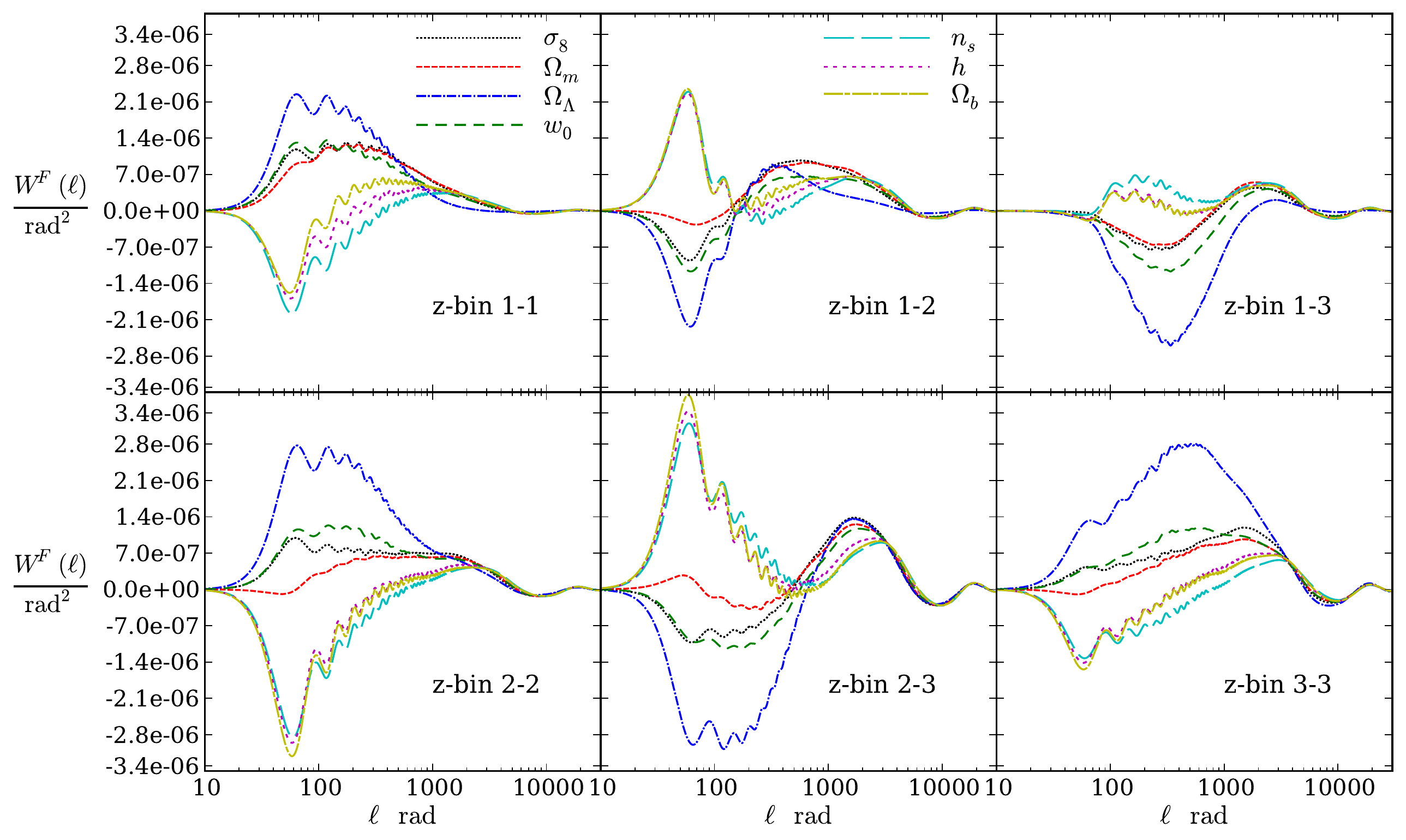}}
    \end{tabular}
    \caption{\small{The filter functions $T^{F}_{+\mu ij}(\vartheta)$ (top panel) and $W^{F}_{\mu ij}(\ell)$ (bottom panel)
      are related to their progenitors via \Eqt\eqref{eqWT}. For
      clarity we show $\vartheta T^{F}_{+\mu ij}(\vartheta)$ on  
      a logarithmic $\vartheta$-scale.
      3 redshift bins and 20 COSEBIs filters defined between 
      $\theta_{\rm min}= 1'$ and $\theta_{\rm max}=400'$ are considered.  
      Each parameter has a different filter function for each redshift pair bin ($z$-bin).       
      A comparison between these functions and their progenitors (see \citealt{2010SEK} and \citealt{2012ASS}) 
      show that they have considerably fewer structures and oscillations.
      The filters for each parameter are normalized according to \Eqt\eqref{eqNormalization}. 
      Hence one can compare how effective each $z$-bin pair is for constraining the parameters.} }
    \label{FigTW}
  \end{center}
\end{figure*}

\fig\ref{FigTW} shows the first-order filter functions, $T^{F}_{+\mu i
  j}(\vartheta)$ and $W^{F}_{+\mu i j}(\ell)$, with $1\leq\mu\leq P=7$
for our fiducial cosmology described in \sect\ref{SectCosmology}.
Here we assume 3 redshift bins and seven free fiducial parameters,
using 20 COSEBIs filters defined between $\theta_{\rm min}= 1'$ and
$\theta_{\rm max}=400'$. The redshift bins were chosen such that they
contain an equal number of galaxies. Since the filters are designed to
maximize the information obtained, their shape shows where most of the
information in $\xi_+(\vartheta)$ or $P_{\rm E}(\ell)$ lies. Here we
choose to only show the first-order filters, although later on we use
the first and second order as well as the combination of both to obtain
the figure-of-merit.  The general trend of $T^{\rm F}$ show that there is
more information about all of the parameters in the higher-redshift
bins and on smaller angular scales.

However, each individual parameter shows a different
pattern for each of the redshift pairs.  For example, the real space
filters, $T^{\rm F}$, for $\Omega_\Lambda$ have significantly higher
amplitudes for combinations of redshift bins 2 and 3 compared to
combinations with the lowest redshift bin.  $\sigma_8$ and
$\Omega_{\rm m}$ filters closely follow each other, although in Fourier space,
i.e. $W^{\rm F}$, the differences are more pronounced. A closer look
at the plots shows that, since these curves are not exactly the same and
also evolve with redshift, it is possible to break their degeneracies
which is present in single-redshift studies
\citep[e.g.,][]{vWMR01,HYG02,2003AJ....125.1014J,Hetter07,Kilbinger13}. 
The oscillations of the $W^{\rm F}$ are a real feature of the CCOSEBIs
weights and do not vanish when more COSEBIs are incorporated in
calculating them. 

The tables \ref{Tabb1z} and \ref{Tabb2z} show the elements of the
compression matrix, $\BMb$, for CCOSEBIs. According to \Eqt\eqref{EqBtoE},
each row of $\BMb$ corresponds to the coefficients for making one of the
CCOSEBIs statistics, $E^{\rm c}_{\mu}$, by linearly combining the COSEBIs
$E_n$. The value of the elements of $\BMb$ show how important each
COSEBIs mode is for building a CCOSEBIs mode.  In both tables, the
element values are much smaller for large $n$ compared to smaller $n$.
As a result we can safely (conservatively) take just the first 20 COSEBIs
to build the compressed statistics.

\begin{table*}
 \caption{The elements of the normalized compression matrix in percentage, $100\times\BM_{\mu n}$, for 1 redshift bin and three parameters. 
 The first column shows the subscript of $E^{\rm c}_\mu$, where one parameter subscripts belong to the first order statistics, $\vc F$, 
 and double subscripts belong to the second order statistics $\vc S$. The first row shows the value of $n$. }
 \begin{center}
%\caption{Fixed table of parameters}
\footnotesize{\begin{tabular}{@{}c@{}|@{}c@{}|@{}c@{}|@{}c@{}|@{}c@{}|@{}c@{}|@{}c@{}|@{}c@{}|@{}c@{}|@{}c@{}|@{}c@{}|@{}c@{}|@{}c@{}|@{}c@{}|@{}c@{}|@{}c@{}|}
\cline{2-16}
&\,1 	\,&\, 2 \,&\, 3\, &\, 4\, &\, 5 \,& \,6 \,&\,7 \,&\, 8 \,&\, 9\, &\, 10\, &\, 11 \,& \,12 \,&\,13 \,&\, 14 \,&\, 15\, 
\\ \hline
\multicolumn{1}{|@{}c@{}|}{\multirow{1}{*}{$\sigma_8$}}	&\,				43.7	\,&\,	$-$59.2	\,&\,	53.4	\,&\,	$-$36.0	\,&\,	17.2	\,&\,	$-$4.8	\,&\,	0.7	\,&\,	$-$1.9	\,&\,	4.5	\,&\,	$-$5.9	\,&\,	5.8	\,&\,	$-$4.6	\,&\,	2.8	\,&\,	$-$1.3	\,&\,	0.3\,\\ \hline
\multicolumn{1}{|@{}c@{}|}{\multirow{1}{*}{$\Omega_{\rm m}$}}&\,				42.1	\,&\,	$-$57.8	\,&\,	53.5	\,&\,	$-$37.9	\,&\,	20.1	\,&\,	$-$7.2	\,&\,	1.8	\,&\,	$-$1.9	\,&\,	4.2	\,&\,	$-$5.9	\,&\,	6.0	\,&\,	$-$4.8	\,&\,	3.0	\,&\,	$-$1.3	\,&\,	0.3\,\\ \hline
\multicolumn{1}{|@{}c@{}|}{\multirow{1}{*}{$\Omega_\Lambda$}}&\,			48.5	\,&\,	$-$61.4	\,&\,	51.0	\,&\,	$-$31.3	\,&\,	13.4	\,&\,	$-$3.1	\,&\,	0.3	\,&\,	$-$1.8	\,&\,	4.3	\,&\,	$-$5.7	\,&\,	5.6	\,&\,	$-$4.4	\,&\,	2.7	\,&\,	$-$1.2	\,&\,	0.3\,\\ \hline
\multicolumn{1}{|@{}c@{}|}{\multirow{1}{*}{$\sigma_8,\sigma_8$}}&\,			46.2	\,&\,	$-$61.2	\,&\,	52.8	\,&\,	$-$32.5	\,&\,	12.5	\,&\,	$-$1.1	\,&\,	$-$1.0	\,&\,	$-$1.7	\,&\,	4.7	\,&\,	$-$6.0	\,&\,	5.6	\,&\,	$-$4.2	\,&\,	2.5	\,&\,	$-$1.1	\,&\,	0.3\,\\ \hline
\multicolumn{1}{|@{}c@{}|}{\multirow{1}{*}{$\sigma_8,\Omega_{\rm m}$}} &\,			38.2	\,&\,	$-$54.5	\,&\,	53.9	\,&\,	$-$41.9	\,&\,	25.4	\,&\,	$-$11.3	\,&\,	3.7	\,&\,	$-$2.2	\,&\,	3.9	\,&\,	$-$5.8	\,&\,	6.2	\,&\,	$-$5.2	\,&\,	3.3	\,&\,	$-$1.5	\,&\,	0.4\,\\ \hline
\multicolumn{1}{|@{}c@{}|}{\multirow{1}{*}{$\sigma_8,\Omega_\Lambda$}}&\,		44.9	\,&\,	$-$61.0	\,&\,	53.8	\,&\,	$-$33.2	\,&\,	11.8	\,&\,	0.9	\,&\,	$-$3.2	\,&\,	$-$0.3	\,&\,	4.5	\,&\,	$-$6.3	\,&\,	5.8	\,&\,	$-$4.1	\,&\,	2.3	\,&\,	$-$0.9	\,&\,	0.2\,\\ \hline
\multicolumn{1}{|@{}c@{}|}{\multirow{1}{*}{$\Omega_{\rm m},\Omega_{\rm m}$}}&\,			47.6	\,&\,	$-$62.0	\,&\,	52.1	\,&\,	$-$30.8	\,&\,	10.9	\,&\,	$-$0.3	\,&\,	$-$1.2	\,&\,	$-$1.7	\,&\,	4.7	\,&\,	$-$5.8	\,&\,	5.4	\,&\,	$-$4.1	\,&\,	2.5	\,&\,	$-$1.1	\,&\,	0.3\,\\ \hline							
\multicolumn{1}{|@{}c@{}|}{\multirow{1}{*}{$\Omega_{\rm m},\Omega_\Lambda$}}&\,		49.8	\,&\,	$-$63.7	\,&\,	51.1	\,&\,	$-$26.2	\,&\,	4.5	\,&\,	5.2	\,&\,	$-$3.8	\,&\,	$-$1.7	\,&\,	5.4	\,&\,	$-$6.1	\,&\,	4.9	\,&\,	$-$3.3	\,&\,	1.9	\,&\,	$-$0.8	\,&\,	0.2\,\\ \hline							
\multicolumn{1}{|@{}c@{}|}{\multirow{1}{*}{\,$\Omega_\Lambda,\Omega_\Lambda$\,}}&\,	48.9	\,&\,	$-$62.8	\,&\,	51.5	\,&\,	$-$28.7	\,&\,	8.4	\,&\,	1.8	\,&\,	$-$2.3	\,&\,	$-$1.4	\,&\,	4.7	\,&\,	$-$5.9	\,&\,	5.3	\,&\,	$-$3.8	\,&\,	2.3	\,&\,	$-$1.0	\,&\,	0.2\,\\ \hline
\end{tabular}}
\end{center}
\label{Tabb1z}
\end{table*}

\begin{table*}
 \caption{The normalized compression matrix elements in percentage, $100\times\BMb$, for 2 redshift bins and three parameters. 
 The first column shows the subscript of $E^{\rm c}_\mu$, where one parameter subscripts belong to the first order statistics, $\vc F$, 
 and double subscripts belong to the second order statistics $\vc S$. The absolute values of the columns of $\BMb$ for each 
 redshift pair decreases rapidly after n=3.}
 \begin{center}
%\caption{Fixed table of parameters}
\footnotesize{\begin{tabular}{@{}c@{}|@{}c@{}|@{}c@{}|@{}c@{}|@{}c@{}|@{}c@{}|@{}c@{}|@{}c@{}|@{}c@{}|@{}c@{}|@{}c@{}|@{}c@{}|@{}c@{}|@{}c@{}|@{}c@{}|@{}c@{}|@{}c@{}|@{}c@{}|@{}c@{}|}
\cline{2-19}
& \multicolumn{6}{|@{}c@{}|}{\,z-bin 1-1\!} & \multicolumn{6}{|@{}c@{}|}{\!z-bin 1-2\,} & \multicolumn{6}{|@{}c@{}|}{z-bin 2-2}
\\ \cline{2-19}
&\,n=1 	\,&\, n=2 \,	&\, n=3\, &\, n=4\, &\, n=5 \,& \,n=6 \,&\,n=1 	\,&\, n=2 \,	&\, n=3\, &\, n=4\, &\, n=5 \,& \,n=6 \,&\,n=1 	\,&\, n=2 \,	&\, n=3\, &\, n=4\, &\, n=5 \,& \,n=6 \,\\ \hline
\multicolumn{1}{|@{}c@{}|}{\multirow{1}{*}{$\sigma_8$}}	& \,	17.8	\, & \,	$-$17.4	\, & \,	10.4	\, & \,	$-$4.4	\, & \,	1.3	\, & \,	$-$0.2	\, & \,	22.2	\, & \,	$-$33.3	\, & \,	31.3	\, & \,	$-$20.3	\, & \,	8.4	\, & \,	$-$1.8	\, & \,	36.8	\, & \,	$-$48.2	\, & \,	41.4	\, & \,	$-$25.8	\, & \,	10.8	\, & \,	$-$2.3\,	\\ \hline
\multicolumn{1}{|@{}c@{}|}{\multirow{1}{*}{$\Omega_{\rm m}$}}	& \,	19.4	\, & \,	$-$19.9	\, & \,	12.8	\, & \,	$-$6.1	\, & \,	2.2	\, & \,	$-$0.4	\, & \,	25.7	\, & \,	$-$37.9	\, & \,	35.7	\, & \,	$-$23.5	\, & \,	10.1	\, & \,	$-$2.2	\, & \,	33.2	\, & \,	$-$43.4	\, & \,	37.1	\, & \,	$-$23	\, & \,	9.6	\, & \,	$-$2.1\,	\\ \hline
\multicolumn{1}{|@{}c@{}|}{\multirow{1}{*}{$\Omega_\Lambda$}}	& \,	13.1	\, & \,	$-$8.2	\, & \,	1.7	\, & \,	0.2	\, & \,	0.1	\, & \,	$-$0.2	\, & \,	1.0	\, & \,	$-$13.6	\, & \,	20.4	\, & \,	$-$16.2	\, & \,	7.3	\, & \,	$-$1.6	\, & \,	50.7	\, & \,	$-$59.7	\, & \,	45.0	\, & \,	$-$24.3	\, & \,	8.9	\, & \,	$-$1.7\,	\\ \hline
\multicolumn{1}{|@{}c@{}|}{\multirow{1}{*}{$\sigma_8,\sigma_8$}}	& \,	12.1	\, & \,	$-$10.6	\, & \,	4.5	\, & \,	$-$0.5	\, & \,	$-$0.5	\, & \,	0.2	\, & \,	13.3	\, & \,	$-$20.8	\, & \,	20.5	\, & \,	$-$14.1	\, & \,	6.2	\, & \,	$-$1.4	\, & \,	43.1	\, & \,	$-$56.7	\, & \,	48.5	\, & \,	$-$29.6	\, & \,	12.1	\, & \,	$-$2.5\,	\\ \hline
\multicolumn{1}{|@{}c@{}|}{\multirow{1}{*}{$\sigma_8,\Omega_{\rm m}$}}	& \,	24.9	\, & \,	$-$26.2	\, & \,	17.9	\, & \,	$-$9.4	\, & \,	3.7	\, & \,	$-$0.8	\, & \,	37.4	\, & \,	$-$51.4	\, & \,	44.5	\, & \,	$-$26.5	\, & \,	10.0	\, & \,	$-$1.8	\, & \,	15.0	\, & \,	$-$21.6	\, & \,	21.1	\, & \,	$-$15.3	\, & \,	7.5	\, & \,	$-$1.9\,	\\ \hline
\multicolumn{1}{|@{}c@{}|}{\multirow{1}{*}{$\sigma_8,\Omega_\Lambda$}}	& \,	4.3	\, & \,	$-$4.0	\, & \,	1.1	\, & \,	1.1	\, & \,	$-$1.2	\, & \,	0.4	\, & \,	5.0	\, & \,	$-$8.0	\, & \,	10.2	\, & \,	$-$9.5	\, & \,	5.7	\, & \,	$-$1.7	\, & \,	44.9	\, & \,	$-$60.7	\, & \,	52.5	\, & \,	$-$31.9	\, & \,	12.7	\, & \,	$-$2.5\,	\\ \hline
\multicolumn{1}{|@{}c@{}|}{\multirow{1}{*}{$\Omega_{\rm m},\Omega_{\rm m}$}}	& \,	13.6	\, & \,	$-$12.8	\, & \,	6.3	\, & \,	$-$1.3	\, & \,	$-$0.3	\, & \,	0.2	\, & \,	17.7	\, & \,	$-$25.2	\, & \,	23.5	\, & \,	$-$15.7	\, & \,	6.9	\, & \,	$-$1.6	\, & \,	41.7	\, & \,	$-$54.8	\, & \,	46.4	\, & \,	$-$28	\, & \,	11.2	\, & \,	$-$2.3\,	\\ \hline
\multicolumn{1}{|@{}c@{}|}{\multirow{1}{*}{$\Omega_{\rm m},\Omega_\Lambda$}}	& \,	9.4	\, & \,	$-$9.5	\, & \,	5.1	\, & \,	$-$1.1	\, & \,	$-$0.4	\, & \,	0.3	\, & \,	13.7	\, & \,	$-$19.4	\, & \,	18.1	\, & \,	$-$12.3	\, & \,	5.6	\, & \,	$-$1.3	\, & \,	44.1	\, & \,	$-$57.9	\, & \,	49.1	\, & \,	$-$29.5	\, & \,	11.6	\, & \,	$-$2.3\,	\\ \hline
\multicolumn{1}{|@{}c@{}|}{\multirow{1}{*}{\,$\Omega_\Lambda,\Omega_\Lambda$}}	& \,	9.5	\, & \,	$-$7.5	\, & \,	2.1	\, & \,	1.0	\, & \,	$-$1.1	\, & \,	0.3	\, & \,	7.8	\, & \,	$-$14.2	\, & \,	15.8	\, & \,	$-$11.9	\, & \,	5.7	\, & \,	$-$1.4	\, & \,	46.5	\, & \,	$-$59.8	\, & \,	49.4	\, & \,	$-$29.1	\, & \,	11.3	\, & \,	$-$2.2\,	\\ \hline
\end{tabular}}
\end{center}
\label{Tabb2z}
\end{table*}

\subsection{Fisher analysis}
\label{subsecFisher}

The Fisher matrix depends on the data, $E^{\rm c}_\mu$, via
\begin{align}
\label{FisherTeg}
\tens{F}_{\mu\nu}= \frac{1}{2} \mathrm{Tr}[\rund{\CMb^{\rm c}}^{-1}\:\CMb^{\rm c}_{,\mu}\:\rund{\CMb^{\rm c}}^{-1}\:\CMb^{\rm c}_{,\nu}
+\rund{\CMb^{\rm c}}^{-1}\:\boldsymbol{\tens {M}}_{\mu\nu}]\;,
\end{align} 
where $\CMb^{\rm c}$ is the data covariance,
$\boldsymbol{\tens{M}}_{\mu\nu}=\vc E^{\rm c}_{,\mu}\; (\vc E^{\rm c}_{,\nu})^\mathrm{t}
+\vc E^{\rm c}_{,\nu}\:(\vc E^{\rm c}_{,\mu})^\mathrm{t}$ and
the comas followed by subscripts indicate partial derivatives with respect to the
cosmological parameters (see \citealt{1997ApJ...480...22T} for
example). We use the same figure-of-merit, $f$, which gives a measure of the mean error on 
parameters,
as in \cite{2012ASS} ,
\begin{equation}
\label{f}
f=\left(\frac{1}{\sqrt{\mathrm{det}\: {\boldsymbol{\tens F}}}}\right)^{1/P}\;,
\end{equation}
where $P$ is the number of free parameters considered.
Furthermore, we have shown in \cite{2012ASS} that for a sufficiently large survey one can
neglect the first term in \Eqt$\eqref{FisherTeg}$ since it does not
depend on the survey area (see \Eqt\ref{EqCov}), while the second term
is proportional to the survey area. Therefore, we neglect the first
term in this study.

% Here we point out that the covariance matrix of the first order compressed statistics 
% is equal to the inverse of the Fisher matrix if an exact COSEBIs covariance is 
% considered in estimating it,
% %
% \begin{equation}
%  \boldsymbol{\tens{F}}=\frac{1}{2} \rund{\boldsymbol{\tens{D}}^{\rm F}}^t \rund{\boldsymbol{\tens{C}^{\rm F}}}^{-1}\boldsymbol{\tens{D}}^{\rm F}
%  =\boldsymbol{\tens{D}}\boldsymbol{\tens{C}}^{-1}\boldsymbol{\tens{D}}^t\boldsymbol{\tens{D}}^{-1}\boldsymbol{\tens{C}}^{-1}
% \end{equation}

\begin{figure}[h]
  \begin{center}
    \begin{tabular}{c}
      \resizebox{85mm}{!}{\includegraphics{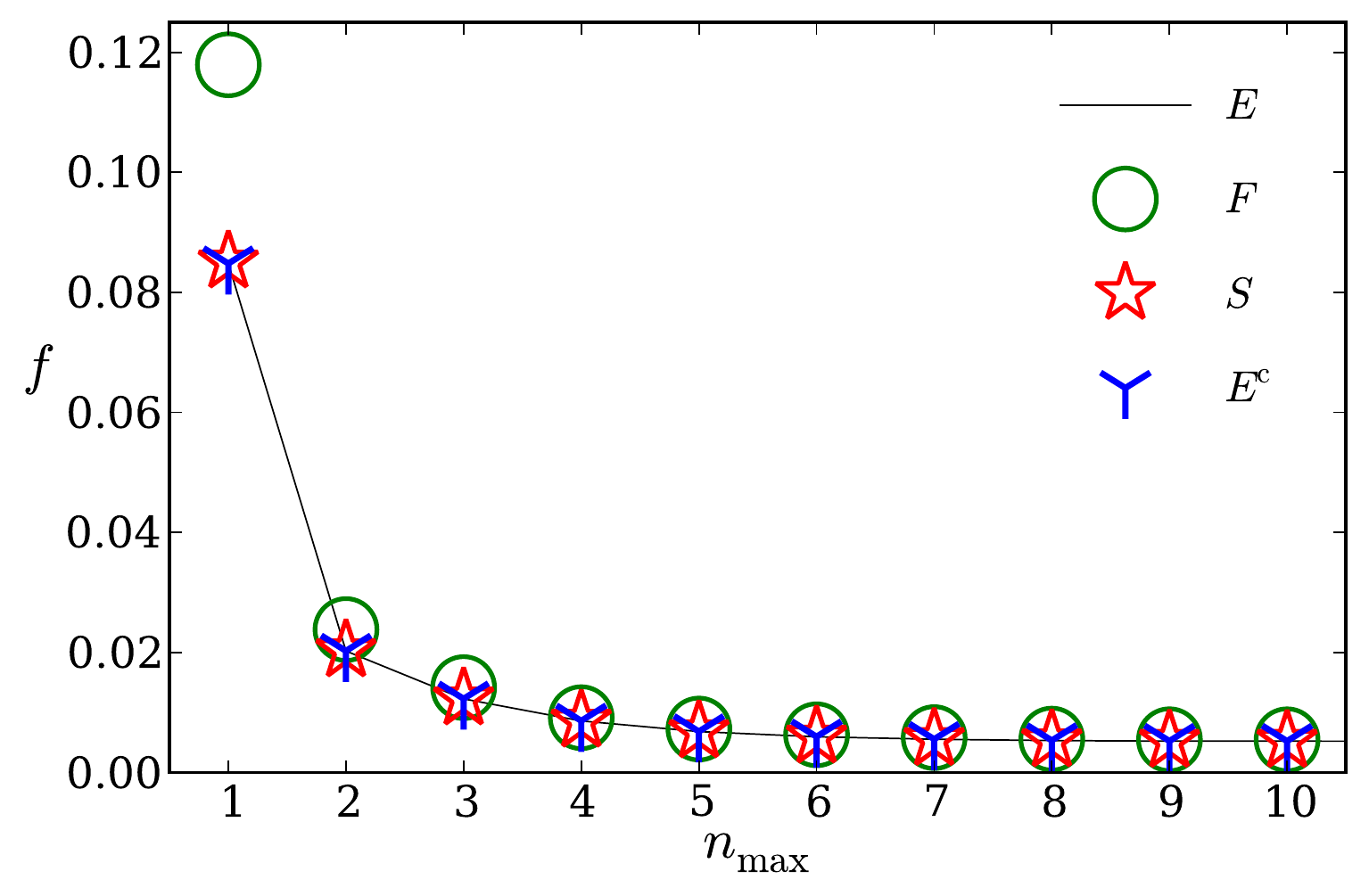}}
    \end{tabular}
    \caption{\small{The figure-of-merit, $f$, as a function of the
        number of COSEBIs, $n_{\rm max}$, used.  7 free parameters listed in
        \tab\ref{Tabcosmparam}, and 8 tomographic redshift bins are
        considered here.  The solid line shows the result for using
        Log-COSEBIs with the true underlying cosmology.  It also
        represents the maximum information level for a given $n_{\rm max}$.  The
        circles, stars and the Y-shaped symbols represent the $f$-values for
        First order, Second order, and their combination $\vc E^{\rm c}$, where
        $n_{\rm max}$ COSEBIs modes with the fiducial cosmological parameters
        are utilized in making them. } }
    \label{FigfNMAX}
  \end{center}
\end{figure}

\fig\ref{FigfNMAX} shows the dependence of $f$ on the number of
COSEBIs modes, $n_{\rm max}$, for 8 redshift bins and 7 free
parameters.  The constrains get tighter as $n_{\rm max}$ increases and
reach a saturation level for all cases. The solid curve shows how much
information can be gained if the $8 n_{\rm max}\times 9/2=36 n_{\rm
  max}$ COSEBIs are used, i.e., the maximum information, or minimum
$f$ value. The points show the amount of information in the first-
($\vc F$) and second-order ($\vc S$) CCOSEBIs, as well as the
combination of both, denoted by $\vc E^{\rm c}$ as before. The
parameters used for 
calculating the covariance matrix to build $\vc F$, $\vc S$ and $\vc
E^{\rm c}$ are that of the fiducial cosmology which are slightly different
from the assumed true parameters (see \tab\ref{Tabcosmparam}).  Nevertheless,
the first-order CCOSEBIs are sufficient to reach a similar Fisher
information level. Notice here that the $\vc F$ statistics for this
case have 7 components,  $\vc S$ have 28 components, and $\vc E^{\rm c}$
have 35, while
for $n_{\rm max}=10$ the COSEBIs have 360 components, i.e., there is at least
an order of magnitude 
difference between the number of statistics for CCOSEBIs and
COSEBIs. Hence, we can obtain the same accuracy of derived parameters
with a highly significant reduction of observables.

The strong reduction of observables needed to cover all the
cosmological information is of great interest with regards to
obtaining accurate covariances, and thus reliable confidence regions
for cosmological parameters. Whereas analytical methods may be able to
obtain approximate covariances \citep[see, e.g.,][and references
therein]{2009MNRAS.395.2065T,2009ApJ...701..945S,Pielorz10,HHS11,2011ApJ...726....7T,2014arXiv1405.2666T}, an
accurate covariance accounting for the complex survey geometry will
probably require extensive simulations. Obtaining the covariance as
sample variance from independent realizations of the simulated
cosmology requires a number of realizations which is about
proportional to the number of observables \citep{2007A&A...464..399H}.
Hence, even a modest reduction in the number of relevant observables
is useful. As we have seen above, the CCOSEBIs serve this purpose very
well.

\begin{figure}
  \begin{center}
    \begin{tabular}{c}
      \resizebox{85mm}{!}{\includegraphics{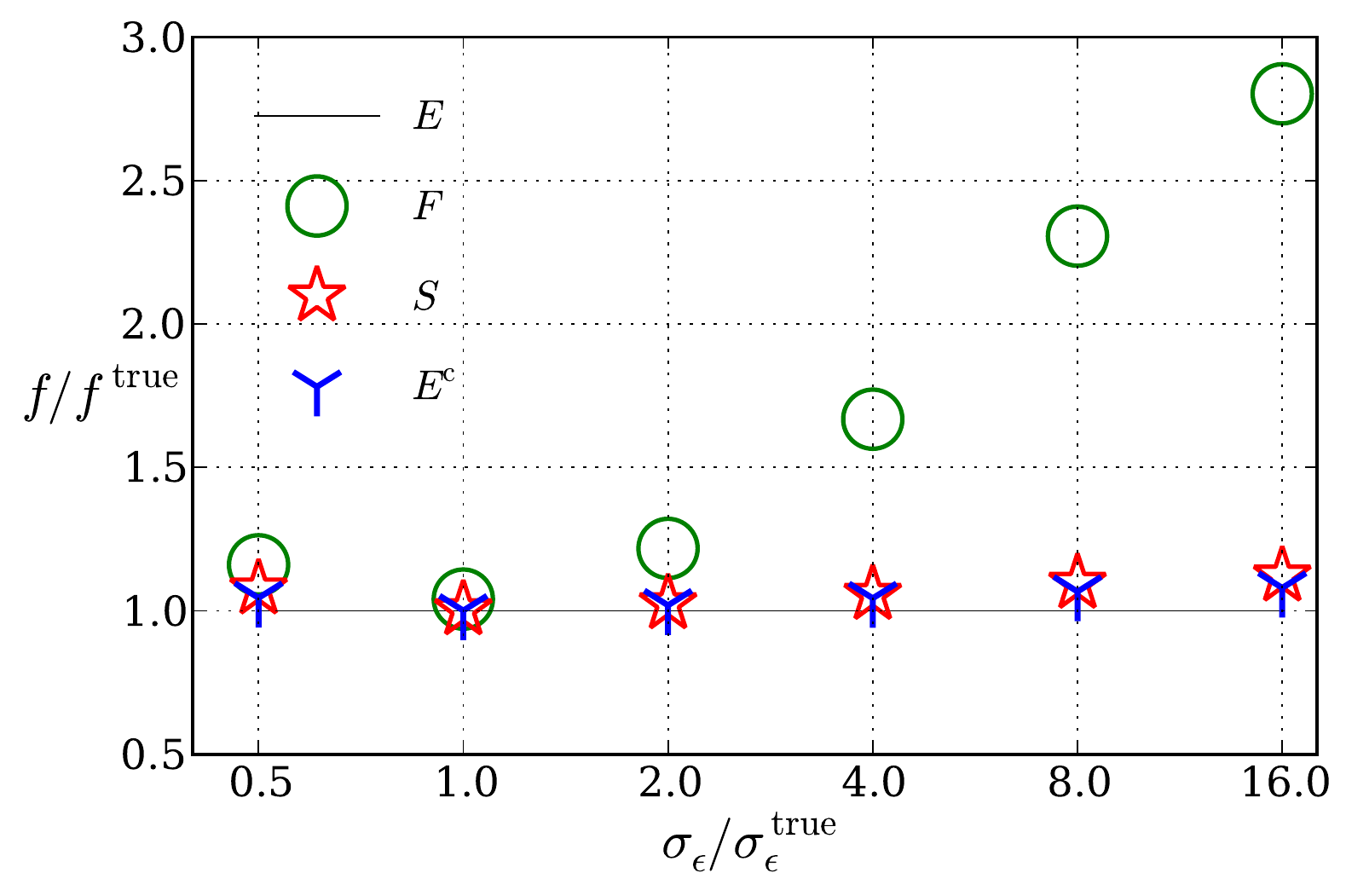}}
    \end{tabular}
    \caption{\small{The figure-of-merit, $f$, as a function of
        $\sigma_\epsilon$.  $f$ is normalized by its minimum value
        which corresponds to using COSEBIs with the correct covariance
        (the solid line).  The intrinsic ellipticity dispersion of
        galaxies, $\sigma_\epsilon$, is varied with respect to its
        true value, 0.3, to show the effects of using a wrong
        covariance.  The markers show the value of $f$ for first
        order, $\vc F$, second order, $\vc S$ and the combination of both $\vc E^{\rm c}$
        CCOSEBIs.}}
    \label{FigSigma}
  \end{center}
\end{figure}

Whereas the construction of the CCOSEBIs requires information about
the covariance, this does not have to be very accurate. In order
to show how using a substantially wrong covariance in defining the
compressed data 
vector impacts on the constrains, we artificially change the value of
$\sigma_{\epsilon}$ which affects the diagonals of the covariance
matrix according to Eqs.\thinspace\eqref{EqCov} and \eqref{EqP}.  \fig\ref{FigSigma} shows $f$
for 7 free parameters, 5 redshift bins and 20 COSEBIs modes, as a
function of the change in the parameter $\sigma_\epsilon$.  $f$ is
normalized by its minimum value, i.e., using COSEBIs with their true
covariance, while $\sigma_\epsilon$ is normalized by its true
value. The first-order statistics, $\vc F$, which has the same dimension
as the parameter space, rapidly diverges from the true Fisher
information limit, while the second order, $\vc S$, and $\vc E^{\rm
  c}$, which span a
larger dimensional space, are much less sensitive to the errors of the COSEBIs
covariance, used for constructing the CCOSEBIs.  Even for a 16 times
larger $\sigma_\epsilon$ the fractional difference between the optimal
$f$ and the measured one from $\vc E^{\rm c}$ is small. Hence, the
consideration of the second-order statistics indeed provides a
powerful mitigation for inaccurate covariances.

Hence we conclude that the method proposed here --
  constructing new observables using an approximate covariance, and
  employing these for cosmological parameter studies -- yields a very
  promising tool for an effective reduction in the necessary efforts
  for constructing accurate covariances. This data compression will
  also be of great help if the covariances are to be obtained from the
  data themselves, e.g., by subdividing the survey region,
  calculating the sample variance on each sub-survey, and scaling the
  result with the ratio of sub-survey to survey area.

\section{Band power}
\label{SectBandPower}
As mentioned in \sect\ref{sectAppCOSEBIs}, any filter function defined
on a finite angular interval which satisfies the constraints
\eqref{EqTCond} can be expressed in terms of the COSEBIs filters.  A
particular filter one might be interested in is a top-hat function in
Fourier space, corresponding to a band power
\citep[e.g.,][]{2003MNRAS.341..100B,2011MNRAS.412...65H}. In this
section we will study how well band powers can be approximated from
correlation functions measured on a finite interval with clean
E-/B-mode separation.

Thus, let $\hat{W}(\ell)$ be a target filter function in Fourier
space, and let us design a filter that approximates $\hat{W}(\ell)$ as
closely as possible. That means we want to find a filter which
minimizes
\begin{equation}
\label{eqDelta}
 \Delta=\int\RMd \ell\: \ell\;\eck{W(\ell)-\hat{W}(\ell)}^2\;,
\end{equation}
where $W(\ell)$ is a linear combination of the $W_n(\ell)$,
\begin{equation}
 W(\ell)=\sum_n c_n W_n(\ell)\;. 
\end{equation}
Note here that the $\Delta$ integral can be defined with a different
weighting of $\ell$, e.g., one can replace $\RMd \ell\: \ell$ in
\Eqt\eqref{eqDelta} with $\RMd \ln\ell$ or simply $\RMd \ell$.  Doing
so does not affect the final estimation accuracy of $\hat{W}(\ell)$
significantly, but may be numerically advantageous.

We want to find the coefficients $c_n$ that minimize $\Delta$;
setting the derivatives of $\Delta$ with respect to $c_m$ to zero, we find
\begin{equation}
\label{EqWW}
 \sum_n c_n \int \RMd \ell\: \ell\; W_n(\ell) W_m(\ell)
=\int\RMd \ell\: \ell\; W_m(\ell)\hat{W}(\ell)\;.
\end{equation}
By defining the matrix
\begin{equation}
 \tens{Q}_{nm}\equiv\int \RMd \ell\: \ell\; W_n(\ell) W_m(\ell) 
\end{equation}
and the vector
\begin{equation}
 V_n\equiv\int \RMd \ell\: \ell\; W_n(\ell) \hat{W}(\ell)\;,
\end{equation}
we can rewrite \Eqt\eqref{EqWW} in matrix form,
\begin{equation}
\boldsymbol{\tens Q} \vc{c}={\vc V}\; \Longrightarrow\; \vc{c}=\boldsymbol{\tens {Q}}^{-1} {\vc V}\;.
\end{equation}
The minimum of $\Delta$ for this solution is 
\begin{equation}
\label{EqDeltamin}
 \Delta_{\rm min}=\int \RMd \ell\: \ell\; \hat{W}^2(\ell)-\vc{V}\boldsymbol{\tens{Q}}^{-1}\vc{V}\;,
\end{equation}
and we quantify the relative deviation of the closest filter $W$ to
$\hat W$ by
\begin{equation}
\label{Eqdeltamin}
\delta_{\rm min}=\Delta_{\rm min}\eck{\int \RMd \ell\: \ell\;
  \hat{W}^2(\ell)}^{-1}\;.
\end{equation}

\begin{figure}
  \begin{center}
    \begin{tabular}{l}
      \resizebox{85mm}{!}{\includegraphics{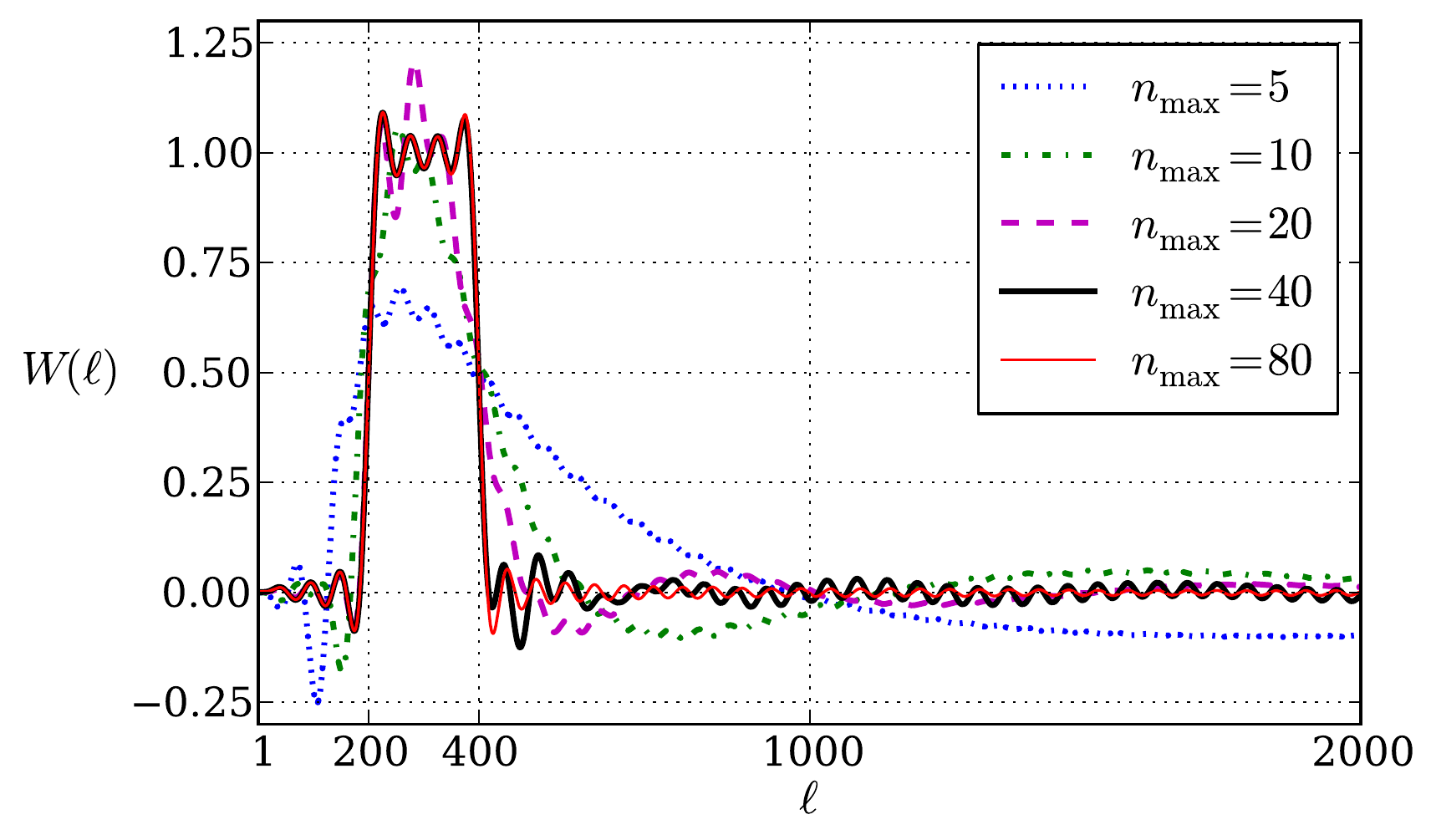}}
    \end{tabular}
    \caption{\small{The estimated top-hat filter function with $\ell_{\rm
min}=200$ and $\ell_{\rm max}=400$, from $n_{\rm max}$ COSEBIs filters
defined on $1'<\vartheta<400'$.  The changes between using 40 and 80
COSEBIs filters are small, so that no better representation is
obtained by using an even higher value of $n_{\rm max}$. }}
    \label{FigWhatN}
  \end{center}
\end{figure}

Filters which satisfy \Eqt\eqref{EqTCond} and vanish outside of the
angular range $[\theta_{\rm min}\,, \theta_{\rm max}]$, are the only
ones that can be represented by COSEBIs.  Hence, applying any filter
that does not satisfy these conditions, on either power spectra or
2PCFs, results in spillage from outside of the measured angular range.
A top-hat function in Fourier space is an example of a filter which is
not well representable by weight functions which correspond to a
finite range in real space.  A top-hat function in Fourier is defined
as $\hat {W}(\ell)=1$ between $\ell_{\min}$ and $\ell_{\rm max}$ and
zero otherwise.  The real space version, $\hat{T}_+(\vartheta)$, of
such a function is
\begin{align}
\label{eqThat}
 \hat{T}_+(\vartheta)&=\int_0^\infty\RMd\ell\,\ell\, \rm J_0(\ell\vartheta) \hat{W}(\ell)\\ \nonumber
 &=\int_{\ell_{\rm min}}^{\ell_{\rm max}}\RMd\ell\,\ell\, \rm J_0(\ell\vartheta)
 =\frac{1}{\vartheta}[\ell_{\rm max}\,\rm J_1(\ell_{\rm max}\vartheta)-\ell_{\rm min}\,\rm J_1(\ell_{\rm min}\vartheta)]\;. 
\end{align}
Using the Parseval's theorem we can find a lower bound for $\Delta$,
\begin{align}
 \Delta&\!=\!\!\int_0^{\infty}\!\!\RMd \ell\,\ell\eck{W(\ell)\!-\!\hat{W}(\ell)}^2\!
 =\!(2\pi)^2\!\int_0^{\infty}\!\! \RMd \vartheta\,\vartheta\,
 \eck{T_+(\vartheta)\!-\!\hat{T}_+(\vartheta)}^2 \nonumber \\ 
 \label{eqLowerBound}
       &\!=\!(2\pi)^2\Bigg\{\int_0^{\theta_{\rm min}}\!\! \RMd \vartheta\, \vartheta\, \hat{T}_+^2(\vartheta)
       +\int_{\theta_{\rm max}}^\infty\!\! \RMd \vartheta\, \vartheta\, \hat{T}_+^2(\vartheta) \\ \nonumber
       &+\!\int_{\theta_{\rm min}}^{\theta_{\rm max}}\!\! \RMd \vartheta\, \vartheta\, \eck{T_+(\vartheta)\!-\!\hat{T}_+(\vartheta)}^2\Bigg\}\;,
\end{align}
where $T_+(\vartheta)$ is the real space form of $W(\ell)$. The sum of the first two integrals in \eqref{eqLowerBound} is the absolute lower
bound on $\Delta$, since the last integral in that equation is non-negative.
Hence, the lower bound for $\delta_{\rm min}$ is
\begin{align}
\label{eqLB}
 \delta_{\rm min}&\geq\delta_{\rm LB}\equiv\!\!\frac{8\pi^2}{\ell_{\rm max}^2-\ell_{\rm min}^2}\Bigg\{
 \int_0^{\theta_{\rm min}}\!\! \RMd \vartheta\, \vartheta\, \hat{T}_+^2(\vartheta)
       +\int_{\theta_{\rm max}}^\infty\!\! \RMd \vartheta\, \vartheta\, \hat{T}_+^2(\vartheta) \Bigg\}\;.
\end{align}
In order to reach the absolute lower bound, the last integral in
\eqref{eqLowerBound} should vanish.  It is necessary and sufficient
for $\hat{T}_+(\vartheta)$ to satisfy the conditions \eqref{EqTCond}
for that to happen, since then $\hat{T}_+(\vartheta)$ can be
represented as a sum over the COSEBIs weights $T_{+n}$.
Inserting the analytic form of
$\hat{T}_+(\vartheta)$ from \Eqt\eqref{eqThat} into
\Eqt\eqref{EqTCond} results in the following two conditions:
\begin{equation}
\label{EqCond1}
\rm I\!:=\!\rm J_0 (\ell_{\rm min}\theta_{\rm max})\!-\!\rm J_0 (\ell_{\rm min}\theta_{\rm min})
\!-\!\rm J_0 (\ell_{\rm max}\theta_{\rm max})\!+\!\rm J_0 (\ell_{\rm max}\theta_{\rm min})\!=\!0\;,
\end{equation}
and
\begin{align}
\label{EqCond2}
 \rm II&:=\theta_{\rm min}^2\rm J_0(\ell_{\rm min}\theta_{\rm min})-\theta_{\rm max}^2\rm J_0(\ell_{\rm min}\theta_{\rm max})\\ \nonumber
 &+\frac{2}{\ell_{\rm min}}[\rm J_1(\ell_{\rm min}\theta_{\rm max})-\rm J_1(\ell_{\rm min}\theta_{\rm min})]\\ \nonumber
 &-\theta_{\rm min}^2\rm J_0(\ell_{\rm max}\theta_{\rm min})+\theta_{\rm max}^2\rm J_0(\ell_{\rm max}\theta_{\rm max})\\ \nonumber
 &-\frac{2}{\ell_{\rm max}}[\rm J_1(\ell_{\rm max}\theta_{\rm max})-\rm J_1(\ell_{\rm max}\theta_{\rm min})]\!=\!0\;,
\end{align}
which should be simultaneously true.  Once the COSEBIs, $E_n$, are
measured from the data, the band power can be estimated by linearly
combining them,
\begin{align}
 \hat{E}&=\frac{1}{2\pi}\int \RMd\ell\:\ell\: \hat{W}(\ell) P_{\rm E}(\ell)\approx\frac{1}{2\pi}\int \RMd\ell\:\ell\: W(\ell) P_{\rm E}(\ell)\\ \nonumber
	&=\sum_n c_n\frac{1}{2\pi}\int \RMd\ell\:\ell\: W_n(\ell) P_{\rm E}(\ell)=\sum_n c_n E_n\;,
\end{align}
\fig\ref{FigWhatN} depicts the convergence to an estimated $W(\ell)$
by increasing the number of COSEBIs modes.  The number of COSEBIs
needed for convergence is substantially higher than the number needed
for constraining parameters with one redshift bin.

\begin{figure}
  \begin{center}
    \begin{tabular}{c}
      \resizebox{85mm}{!}{\includegraphics{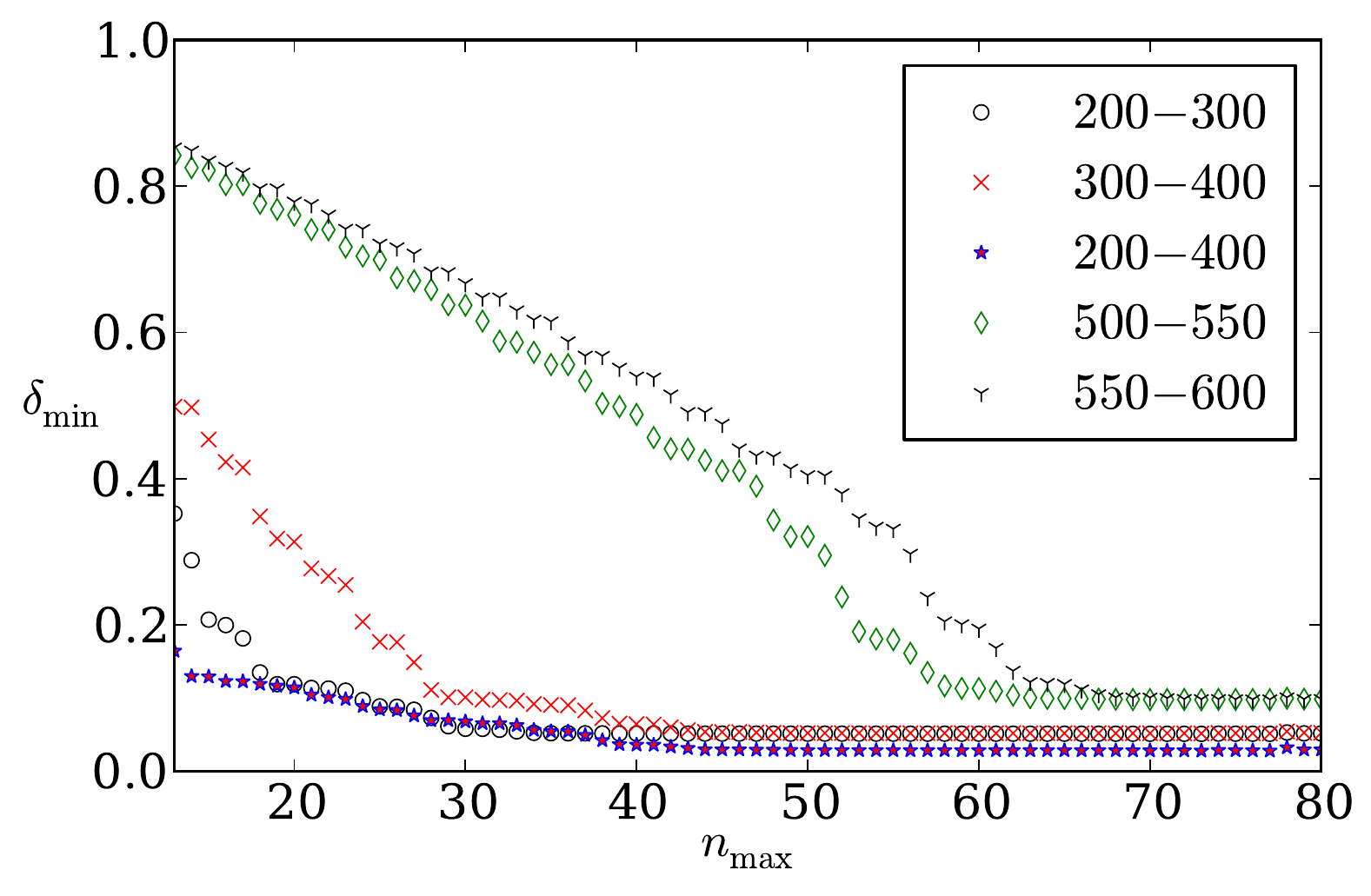}}
    \end{tabular}
    \caption{\small{The relative difference, $\delta_{\rm min}$,
        between the estimated top hat and the input as a function of
        the number of COSEBIs filters, $n_{\rm max}$, utilized for a
        few $\ell$-ranges.  In all cases the saturation level is
        reached before $n_{\rm max}=80$. The minimum value of
        $\delta_{\rm min}$ is shown in \tab\ref{TabEband}. In general,
        a higher number of modes is needed for a narrower band power,
        which is due to the spillage beyond the observed angular range
        (see \Eqt\ref{eqLB}).} }
    \label{FigDeltaMin}
  \end{center}
\end{figure}

\fig\ref{FigDeltaMin} shows the dependence of $\delta_{\rm min}$ on
the number $n_{\rm max}$ of COSEBIs.  Here we demonstrate that for all angular
ranges considered, a saturation level is reached, i.e., adding more
COSEBIs filters will not lead to a smaller difference between the
estimated $W$ and the top hat.  Furthermore, in \tab\ref{TabEband} we
show the value of $\delta_{\rm min}$ for 80 COSEBIs which can be
compared to its lower bound, $\delta_{\rm LB}$ (see \Eqt\ref{eqLB}),
and the relative difference between the estimated band power and its
true value, $\delta_{\rm band}=(\hat{E}-E)/\hat{E}$.  The saturated
$\delta_{\rm min}$ values are larger than but close to $\delta_{\rm LB}$.
The difference between the two arises from violating conditions
\eqref{EqCond1} and \eqref{EqCond2}.  In the table, we use three
$\ell$-weighting schemes, which do not change the $\delta_{\rm min}$
values significantly.  However, the estimated band-power deviations,
$\delta_{\rm band}$, can vary by more than a few percent between the
cases.  This is due to the spillage of the estimated band
power and the fact that the $\ell$-weighting scheme decides which way
the spillage is directed to. The $\delta_{\rm band}$ values are
cosmology dependent and can be very different for a power spectrum
with more features.

It is interesting to note that the deviations $\delta_{\rm band}$ of
the estimated band powers from their true values are in most cases
considerably {\it smaller} than the relative deviation $\delta_{\rm
  min}$ between the top-hat filter and the best representation of the
top hat by COSEBIs weight functions. This, however, is an effect of
the properties of the power spectrum in our assumed cosmological
model: the power spectrum is sufficiently smooth that the spilling
caused by the effective weight $W(\ell)$ out of, and into the range of
the top hat, largely compensate each other \citep[see][for a related
discussion on band powers in cosmic shear analysis]{SvWKM}. Hence, the
fact that $\delta_{\rm band}$ is relatively small is {\it not} a
statement about the accuracy of the method of band-power estimates,
but rather a consequence of the properties of the power
spectrum. But the latter should be probed by estimating the band
power. Thus, it would be strongly misleading to judge the accuracy of
the method on presumed properties that rather ought to be
investigated. Indeed, the quantity $\delta_{\rm min}$ yields an
estimate on the accuracy with which band powers can be obtained.

\begin{table}
 \caption{Examples of band-power estimation from 80 COSEBIs for $[1',400']$. 
 The first column shows the $\ell$-range of the top-hat function $\hat{W}(\ell)$. 
 The rest of the columns show percentage values for minimum relative difference between the estimated and the top hat, $\delta_{\rm min}$, 
 the absolute minimum (Lower Bound) value for $\delta_{\rm min}$, $\delta_{\rm LB}$ (see \Eqt\ref{eqLB}), 
 and the relative difference between the estimated band power and its true value, $\delta_{\rm band}=(\hat{E}-E)/\hat{E}$, for different 
 $\ell$ weightings, respectively. $\delta_{\rm min}$ and $\delta_{\rm LB}$ values correspond to the $\RMd\ell\,\ell$ weighting. The values 
 of these quantities for other cases are similar. 
 The cosmological model used here is the fiducial model from \tab\ref{Tabcosmparam}, with one redshift bin.}
 \begin{center}
%\caption{Fixed table of parameters}
\footnotesize{\begin{tabular}{|c|c|c|c|c|c|}
\hline
{\,$l_{\rm min}-l_{\rm max}$} &$\delta_{\rm min}$  &$\delta_{\rm LB}$   &$\delta_{\rm band},\RMd \ell\,\ell$ &  $\delta_{\rm band},\RMd \ell$ & $\delta_{\rm band},\RMd \ell/\ell$    \\ \hline
{\,$200-300$\,}                 &5.18     	   &5.18	        &3.31  				   &0.72			    &2.29 \\ \hline
{\,$300-400$\,}   		&5.27  		   &5.24	        &6.78				   &1.74			    &2.99\\ \hline
{\,$200-400$\,}   		&2.95 		   &2.92	        &4.94				   &1.20			    &2.62\\ \hline	
{\,$500-550$\,}   		&9.86 		   &9.85	        &10.24				   &3.03			    &4.27\\ \hline	
{\,$550-600$\,}   		&9.88 		   &9.87	        &14.90				   &7.83			    &9.14\\ \hline
\end{tabular}}
\end{center}
\label{TabEband}
\end{table}

\section{Summary and Discussion}
\label{SectConclusions}

Data compression is an important challenge to tackle for future
cosmological surveys.  It is essential for estimating accurate
covariances.  Current cosmological surveys such as Planck provide us
with tight constraints on most cosmological parameters.  This motivated
us to define combinations of statistics inspired by their low-order
Taylor expansion around a fiducial cosmological model.  The strategy
for finding the compressed statistics involves first- and second-order
derivatives of a parent statistics with respect to the free parameters
as well as their covariance.  The statistics corresponding to the
first order derivatives, $\vc F$, have the same dimension as the
parameter space, while the statistics derived from second-order
derivatives, $\vc S$, provide a possibility to span a
larger-dimensional space.
Consequently, $\vc F$ is more sensitive to the choice of the fiducial
cosmology and covariance.  The combination of $\vc F$ and $\vc S$,
enables one to use well-defined and motivated sets of statistics which
alleviate many of the data analysis problems.  In total the number of
compressed statistics is $P(P+3)/2$, where $P$ is the number of free
parameters in the model.

In the case of a cosmic shear analysis, the COSEBIs already provide an
effective compression compared to other two-point statistics, e.g.,
the shear two-point correlation functions. However, adding tomographic
bins, which is necessary for intrinsic alignment corrections,
substantially increases the number of observables.  As a result,
further data compression is required. We applied our compression
formalism to Log-COSEBIs to study its properties.  We found that for a
well-estimated COSEBIs covariance matrix, the first-order compressed
statistics are sufficient.  However, as mentioned above, the accuracy
of covariance estimations from simulations depend on the number of
observables incorporated. The higher this number is, the more
simulations are needed which rapidly becomes too
expensive. Consequently, we used highly inaccurate covariances for
defining the compressed COSEBIs (CCOSEBIs), to test their efficiency
for such cases. We found that the figure-of-merit obtained from the
first-order CCOSEBIs deviates substantially from the optimal
information level as the difference between the assumed COSEBIs
covariance and their true covariance increases. In contrast, the set
of second-order CCOSEBIs is far less sensitive to the choice of
covariance, owing to its larger dimensionality. The combination of
both is basically insensitive to the accuracy of the covariance, at
least in the framework of the simple model that we have tested here.
Consequently, we propose that this strategy is applicable for the
future data analysis. We note that our first-order CCOSEBIs is
equivalent to the Karhunen--Lo\`eve data compression (with
parameter-independent covariance) in the that the covariance is
accurately known \citep{1997ApJ...480...22T}.

In this paper we used a Fisher analysis, which assumes the parameters
have a normal distribution, to compare the constrains from COSEBIs and
CCOSEBIs.  Both Fisher matrix and $\vc F$, the first order compressed
statistics depend only on the first order derivatives and the covariance. 
If the fiducial cosmology coincides with the truth and the
covariance is exact, then the $\vc F$ is equivalent to a
Fisher formalism, since in this case the derivative matrix of $\vc F$ is 
equal to its covariance matrix which is consequently equal to the Fisher
matrix. However, when the covariance deviates from the truth the differences become visible. 
For our future studies we plan to use likelihood analysis which does not
make assumptions about the Gaussianity of the 
likelihood with respect to the model parameters.

The COSEBIs filter functions form a complete basis for any filter that
satisfies \Eqt\eqref{EqTCond} which are necessary and sufficient
conditions for a clean E-/B-separation on a finite interval, together
with the condition that the
filters should also vanish outside of the finite angular range.
Consequently, any filter that satisfies these conditions can be
represented by a linear combination of the COSEBIs filters.  In this
paper we showed how any given weight function can be mimicked by
COSEBIs weights. In particular, we tried to represent top-hat
filters in Fourier space using this strategy.  We found that, due to
the infinite support of a Fourier top-hat in real space, an accurate
representations of them is impossible.  This task becomes harder as
the top hat and the angular range get narrower.  Consequently, band
convergence power spectra estimated from finite angular range
information will suffer from spillage, hence they will be
inaccurate and biased, in a way that is dependent on the power
spectrum -- the quantity to be probed.  Hence, we
caution against using narrow-band power spectra for cosmic shear
analysis. The estimated  powers are relatively accurate if the power
spectra are rather smooth functions of $\ell$. However, for such
smooth functions, there are better ways to characterize them than
using band powers, such as presenting them by a set of basis
functions.  We thus see
no advantage in using power spectra for cosmic shear analysis
on a finite angular range.

\begin{acknowledgements}
  We thank  Andy Taylor for interesting discussions.
  This work was supported in part by
the Deutsche Forschungsgemeinschaft under the TR33 `The Dark
Universe'.
\end{acknowledgements}

\bibliographystyle{aa}
\bibliography{COSEBIs}

% \begin{thebibliography}{}
% 
% \end{thebibliography}

\end{document}